\documentclass[prd, aps, superscriptaddress, twocolumn, preprintnumbers, floatfix, nofootinbib]{revtex4}
\usepackage[dvips]{graphicx}
\usepackage{epsf}
\usepackage{amsmath}
\usepackage{amssymb}

\usepackage{graphicx}
\usepackage{dcolumn}
\usepackage{bm}

\def\be{\begin{equation}}
\def\ee{\end{equation}}
\def\bea{\begin{eqnarray}}
\def\eea{\end{eqnarray}}

\usepackage{color}

\usepackage[unicode=true,pdfusetitle,
 bookmarks=true,bookmarksnumbered=false,bookmarksopen=false,
 breaklinks=false,pdfborder={0 0 1},backref=false,colorlinks=true]
 {hyperref}
\hypersetup{
 linkcolor=blue, citecolor=magenta, urlcolor=red, filecolor=blue}

\begin{document}

\title{Varying fundamental \emph{constants}: a full covariant approach and
cosmological applications}
\author{Guilherme Franzmann\\
{\small{} \textit{Department of Physics, McGill University, Montreal, QC, H3A
2T8, Canada}}\\
{\small{}email: \href{mailto:guilherme.franzmann@mail.mcgill.ca}{guilherme.franzmann@mail.mcgill.ca}}}

\begin{abstract}

We build a minimal extension of General Relativity in which Newton's
gravitational coupling, $G,$ the speed of light, $c$, and the cosmological
constant, $\Lambda,$ are spacetime variables. This is done while
satisfying the contracted Bianchi identity as well as the local conservation
of energy momentum tensor. A dynamical constraint is derived, which
shows that variations of $G$ and $c$ are coupled to the local matter-energy
physical content, while variation of $\Lambda$ is coupled to the
local geometry. This constraint presents a natural cosmological screening
mechanism that brings new perspective concerning the current observations
of a cosmological constant, $\Lambda_{0},$ in cosmological observations.
We also explore early universe background cosmology and show that
the proposal provides alternatives to obtain an accelerated expansion,
similar to those coming from Varying Speed of Light theories. 
\end{abstract}

\maketitle

\section{INTRODUCTION}

Fundamental constants have played an important role in physics since
their very first appearance in Newton's theory of gravity, where he
introduced the gravitational constant, $G.$ Usually, they are directly
connected to the strength of a particular interaction that happens
in nature, but also can be used to define regimes for which a particular
theory will remain valid, as it is the case for Planck's constant,
$h,$ and the speed of light\footnote{Throughout this paper, we will refer to $c$ as the speed of light,
which is the common terminology when talking about the variation of
fundamental constants. However, for this work, one should actually
think of it as the ``spacetime speed'', in the sense discussed by
Ellis and Uzan \cite{EllisUzan}. In light of their work, here we
are considering the spacetime speed, the speed that appears in the
metric, and the Einstein's speed, the speed that appears in Einstein's
equations, to be the same, \emph{i.e.}, $c_{ST}=c_{E}=c\left(x^{\alpha}\right)$,
and being spacetime variables.}, $c.$ 

Looking at the history of physics, we have seen different examples
of dimension-ful and -less constants that ended up not being constants
after all. An example of the former is the acceleration due to gravity
on the Earth's surface, $g\simeq9.8\,\text{m/\ensuremath{s^{2}}},$
which was thought to be a constant before Newtonian gravity, while
an example of the latter is the fine structure constant, $\alpha\equiv e^{2}/4\pi\varepsilon_{0}\hbar c,$
$e$ is the electric charge and $\varepsilon_{0}$ is permittivity
of free space, which is actually a function of the energy scale being
considered. 

Therefore, it is not a surprise that even before the 1900's Kelvin
and Tait \cite{KelvinTait} were already considering variations of
the speed of light. Later, Dirac \cite{Dirac} considered cosmological
variations of $G$, opening room for other approaches that resulted
in the Jordan-Brans-Dicke theories \cite{Jordan, BransDicke, Canuto}.
It is also important to mention that even the electric charge has
been considered to be varying, a proposal first accounted by Bekenstein
\cite{Bekenstein}, after considering an $\alpha$-varying theory.
Therefore, we can already appreciate the relevance of inquiring about
the constancy of the fundamental constants considered today in our
theories. 

Epistemologically, we can observe that a fundamental constant remains
to be so until we figure out a more fundamental model in which the
aforementioned constant becomes actually a variable that assumes a
particular value for the regime so far considered. A very clear example
to illustrate this is the compactification procedure for extra dimensions
in String Theory, which makes the Newtonian coupling dependent on
the moduli fields \cite{Polchisnki}, so $G$ is only effectively
a constant. Another one encompasses a large set of models commonly
referred to Varying Speed of Light (VSL) theories \cite{Magueijo1},
which are generally concerned with variations of the speed of light
in the history of the universe. 

In this work we will be interested in promoting $G,$ $c$ and $\Lambda$,
the cosmological constant, to be functions of the spacetime coordinates.
All these quantities are dimensionful and it is an important matter
to distinguish if their variations will be physical in any sense.
In fact, there has been an extensive debate over the years regarding
the meaning of considering a dimensionful fundamental constant to
be varying. This debate has orbited much more the VSL theories, since
these are the ones that became more popular after solving some of
the early universe puzzles without invoking an inflationary phase
\cite{Magueijo1}. The main topic of the discussion has been the physical
relevance of considering a dimensionful constant to vary, once we
can always choose a different system of units in which this variable
would be a constant again, so that its variation would merely be a
unit system artifact. For people who are more concerned with the \emph{conservative}
side of the question, we recommend \cite{EllisUzan,Duff} and, in
particular, \cite{Ellis}; for the \emph{liberal} counterpart, it
is worth checking \cite{Magueijo1,MagueijoMoffat,Moffat1}. Our position
on this issue will be addressed later. 

Regardless of the \emph{political} orientation on this matter, it
is fundamental to consider what the experimental constraints concerning
variations of the fundamental constants are. The most interesting
result allowing for a significant variation is related to the fine
structure constant. In the work of Webb \emph{et al. }\cite{webb},
they seemed to have found evidence for a slow increase of $\alpha$
in time for redshifts between $0.5$ and $3.5.$ These results have
been questioned and an update on this discussion can be found in \cite{MoffatGraham}.
For the other constants, most of the constraints come from experiments
using atomic clocks, the Oklo phenomenon, Solar System observations,
meteorites dating, quasar absorption spectra, stellar physics, pulsar
timing, the Cosmic Microwave Background (CMB) and big bang nucleosynthesis
(see \cite{uzanreview,barrowparsons }). Although most of these experiments
have left little room for considering variation of the fundamental
constants, they are redshift and spatially constrained \cite{BarrowToole},
not to mention that the interpretation of the results might be changed
if someone had considered a model in which these constants would be
varying as a prior. One thing is certain: if we keep ourselves to
the early universe, varying fundamental constants has not been ruled
out.

The proposal presented here is an extension of General Relativity
(GR) in which $G,$ $c$ and $\Lambda$ are allowed to vary, while
preserving the two fundamental ingredients that were considered by
Einstein: the underlying geometrical structure of the theory, namely
the requirement of satisfying the (contracted) Bianchi identity, and
the local conservation of the energy momentum tensor. It is important
to acknowledge that there is a vast literature concerning models in
which either $G,$ $c$ or $\Lambda$ are varying, almost all of them
at the background level. Some of them have an overlap with the approach
developed here, and we will highlight some of the differences and
similarities. Moreover, it is relevant to say that the model presented
here has already appeared in the literature at the background level
\cite{avelino,belinchon}, but some of the conclusions and discussions
we will draw are original. More importantly, this paper aims to open
room for a new framework that is being developed in which the dynamics
associated to the variation of the fundamental constants is given
by a potential for these variables  \cite{cflation}, which can be seen as scalar fields
in an action. This way we shall address one of the biggest criticisms
towards some of the VSL theories so far: the lack of a well defined
variational principle. 

This paper is divided as follows: in section \ref{sec:um}, we briefly
comment on the previous literature regarding varying fundamental constants
in cosmology focusing mainly in their advantages and drawbacks; in
section \ref{sec:The-proposal} we discuss and provide the framework
we consider throughout this paper; section \ref{sec:discussions}
considers the generalized version of the Friedmann's equations and
a simple application of them; section \ref{sec:tres} presents a sort
of \emph{bootstrap} mechanism that provides a possible explanation
of why we seem to observe a cosmological constant given this framework;
section \ref{sec:Prospects} brings attention to future applications;
finally we conclude in section \ref{sec:Conclusions}.

\section{Varying fundamental constants\emph{ }literature \label{sec:um}}

There is an extensive literature of models that considers variation
of classical (\emph{i.e. }non-quantum) fundamental constants, mainly
$G$ and $c$, as well of the cosmological constant\footnote{Note that although a bare cosmological constant should always come
for free in Einstein's equations, we know that, if the vacuum energy
gravitates, its effect would be given by a cosmological constant in
the energy side of the equations. Since the vacuum energy is a quantum
effect, we do not consider the overall $\Lambda$ in the equations
to be a classical constant. }, $\Lambda$, and $\alpha$. We will briefly discuss some of these
models so that we can make comparisons among them and be aware of
the advantages and drawbacks of each model. Since our objective is
not to review, the reader should not expect an exhaustive list of
them. 

\subsection{$\Lambda\left(t\right)$ models}

There are different motivations to consider the cosmological constant
to be varying. For the very early universe, for example, this could
easily solve the flatness and horizon problems \cite{cflation}. In
fact, to some extent inflation does that in a more elaborated fashion,
since during the slow-roll phase the stress tensor of the scalar field
is given basically by the potential of the field, which acts effectively
as a $\Lambda\left(t\right).$ For the late time universe, this is
also interesting since it could account for backreaction effects coming
from the effective energy-momentum tensor due to fluctuations at second
order Einstein's equations that also influence the background cosmology
(for a review on backreaction, see \cite{backreactionrober}). These
effects also play a role in the early universe (see \emph{e.g.} \cite{backreactionour}). 

Given these motivations, some of the models that have been considered
so far include the time dependence of $\Lambda$ by considering $\Lambda\left(H\left(t\right)\right)$
\cite{lima}, where $H\left(t\right)$ is the Hubble parameter; or
by $\Lambda\left(a\left(t\right)\right)$ \cite{lambdaChen}, $a\left(t\right)$
the scale factor; also by $\Lambda\left(\alpha\left(t\right)\right)$
\cite{lambdalapha}, $\alpha\left(t\right)$ the varying fine structure
constant; and even combination of $H$ and $a$, \cite{lambdaHandA}.
All these models have different and interesting cosmologies, but as
it will be clear later on, if we only assume that $\Lambda$ is varying
without considering also $G$ and/or $c$ to be varying as well, and
having the dynamics solely given by Einstein's equations, thus either
Bianchi identity or the local conservation laws will be violated.

\subsection{Bimetric models }

In general, bimetric models make use of two different metrics: $g_{\mu\nu}$,
the gravitational field, and $\hat{g}_{\mu\nu},$ the metric that
couples to matter. Usually $\hat{g}$ depends on the gravitational
field plus a new scalar \cite{bimetricscalar}, or vector field \cite{bimetricvector}.
This implies that massless particles will have different velocities,
so that special relativity will be realized differently in each of
these sectors. Due to the existence of the two metrics, the coupling
between geometry and matter in Einstein's equations is changed and
picks a dependence on the dynamics of this new field that has been
introduced. Some of the consequences of such approaches are the resolution
of the flatness and horizon problem in a Friedmann-Robertson-Walker
(FRW) universe, also providing a graceful exit to the inflationary
epoch, and even recovering a scale invariant spectrum for the fluctuations
\cite{bimetricmodels}. 

However, their successes are limited since the very existence of a
different metric coupling to the matter sector explicitly violates
the equivalence principle. Moreover, although the Bianchi identity
is satisfied in relation to the metric $g,$ the local conservation
laws are satisfied in relation to the metric $\hat{g}$.

\subsection{Jordan-Brans-Dicke framework}

The very first implementation of a varying $G$ model after Dirac's
initial phenomenological proposal was made by Jordan \cite{Jordan}.
This was an important step, since it incorporated the variation of
$G$ as a dynamical feature of the model instead of just being imposed
by hand, setting up a consistent framework from which equations of
motion could be derived in a consistent matter following an action
principle. The action proposed by him was given by:
\begin{equation}
S=\int d^{4}x\sqrt{-g}\phi^{\eta}\left[R-\xi\left(\frac{\nabla\phi}{\phi}\right)^{2}-\frac{\phi}{2}F^{2}\right],
\end{equation}
where $\eta$ and $\xi$ are two parameters, and $F$ is the electromagnetic
field strength. It then follows that $G$ and $\alpha$ are promoted
to be dynamical variables. 

As it is summarized in \cite{uzanreview}, later it was realized that
if $\eta\neq-1$ the atomic spectra will be space-time dependent.
After fixing $\eta=1$, the model becomes 1-parameter dependent only
and represents a class of scalar-tensor theories in which only $G$
is a dynamical variable. This idea was further explored by Brans and
Dicke \cite{BransDicke}. The most recent work that makes use of this
approach is the BSBM proposal, which actually considers the variation
of $G$ and $\alpha,$ and it can be seen as combination of the initial
proposal from Bekenstein \cite{Bekenstein} revived by Sandvik, Barrow
and Magueijo \cite{SBM}.

\subsection{VSL theories}

Besides the proposals we have briefly overviewed above, there is a
whole class of models typically called VSL theories. There is a big
overlap among them and their consequences, so here we focus in two
seminal papers, Moffat's early work \cite{mofat1} and Albrecht and
Magueijo's paper \cite{albrechtmagueijo}. It is historically fair
to bring attention to the fact that Moffat's first paper was published
in 1993, and his ideas were mostly neglected until the work of Albrech
and Magueijo, which finally got the proper attention to VSL models
from the community.

In summary, the initial idea brought up by Moffat involved a spontaneous
breaking of Lorentz invariance (associated to a first order phase
transition) in the early universe, and this symmetry would be later
restored. The symmetry breaking was produced after introducing a Higgs
mechanism for four scalar fields. The flatness and horizon problems
are solved with the phase transition, which also leads to a scale
invariant power spectrum for the energy density fluctuations. It is
worth mentioning that in order to solve these problems the speed of
light has to drop by $10^{28}$ between the two phases. It preserves
the underlying geometrical structure by satisfying Bianchi identity,
but that leads to non-standard energy-momentum conservation laws. 

Albrecht and Magueijo's proposal also considers departures from Lorentz
invariance, as well as violation of the local conservation laws. The
core idea is to postulate that Friedmann's equations, as known from
General Relativity in a cosmological setting, should remain the same
in the CMB frame, but having now $c$ and $G$ being also functions
of the cosmological time (see discussion in section \ref{sec:discussions}).
This explicitly breaks covariance, and also implies that Einstein's
equations, as we know them, would only be valid in the CMB frame.
Although Bianchi identity is enforced, this also results in non-standard
local conservation laws, which now includes source terms proportional
to $\dot{c}/c$. Within such a prescription, Albrecht and Magueijo
are able to also solve the horizon and flatness problems, as well
as alleviating the cosmological constant problem. Moreover, they can
also account for the large entropy inside the horizon nowadays. Unfortunately,
it lacks a dynamical equation for $c\left(t\right),$ so that one
has to impose its dynamics by hand.\\

For a more extensive review and discussion of different models, please
see \cite{Magueijo1,barrowreview,barrow3}.

\section{The framework of the proposal\label{sec:The-proposal}}

We have seen above that among the different ideas that have been considered
so far regarding the variations of the fundamental constants, most
of them lack a constraint for the variations of the different \emph{constants}
coming from theoretical grounds, which usually leads to violation
of the conservation laws or of the very geometrical structure in which
these theories are considered. Here, we aim to revive an approach
that preserves the underlying geometrical structure and the known
local conservation laws.

The proposal is very simple. We start off by considering the (contracted)
Bianchi identity:
\begin{equation}
\nabla^{\mu}G_{\mu\nu}=0,\label{eq:Bianchi Identity}
\end{equation}
where $G_{\mu\nu}\equiv R_{\mu\nu}-\frac{1}{2}g_{\mu\nu}R$ is known
as the Einstein tensor, and $R_{\mu\nu}$ and $R$ the Ricci tensor
and scalar, respectively. This is a formal identity after assuming
a torsion-free connection and the metricity condition, $\nabla_{\rho}g_{\mu\nu}=0$,
being valid for any (pseudo)-metric manifold \cite{Wald}. We also
consider the minimally coupled local conservation law:
\begin{equation}
\nabla^{\mu}T_{\mu\nu}=0,\label{eq:LocalConservation}
\end{equation}
where $T_{\mu\nu}$ is the energy-momentum tensor (EMT). It is no
surprise for all the readers that are familiar with General Relativity
that those two equations can be written together in a self-consistent
way through Einstein's equations, namely,
\begin{equation}
G_{\mu\nu}=\frac{8\pi G_{0}}{c_{0}^{4}}T_{\mu\nu}+\Lambda_{0}g_{\mu\nu},\label{eq:einstein's equation normal}
\end{equation}
where we have allowed the presence of a cosmological constant\footnote{From now on the subscript 0 denotes the quantity is a constant.},
$\Lambda_{0}.$ Besides, the proportionality constant multiplying
the EMT tensor is recovered after one considers the Newtonian limit
\cite{Wald}. It is important to note that Einstein's equations are
supposed to rule the dynamics of the spacetime structure in the presence
of matter/energy content. Even though the local experiments at the
time did not indicate the existence of any space and/or time variation
concerning the so called fundamental constants, as the speed of light,
$c_{0},$ and the Newton's gravitational constant, $G_{0},$ it was
a conservative call to consider that those quantities would be necessarily
constants for all times and regions of the universe. Therefore, our
proposal is to consider the more general case, in which the constants
are promoted to be spacetime variables, while keeping intact the underlying
geometrical structure of General Relativity, namely, we preserve the
validity of (\ref{eq:Bianchi Identity}) and the local conservation
laws (\ref{eq:LocalConservation}).

In order to do so, let us take a step back and write the most general
relation between Einstein's tensor and the energy-momentum tensor
to be:
\begin{equation}
G^{\mu\nu}=\chi\left(x^{\rho}\right)T^{\mu\nu}+g^{\mu\nu}\Lambda\left(x^{\rho}\right),
\end{equation}
where we have now allowed a coordinate dependence on the cosmological
constant-like term, represented by $(x^{\rho}),$ as well for the
coupling of the EMT, instead of assuming that $\alpha$ and $\Lambda$
are constants, as it was done for equation (\ref{eq:einstein's equation normal}).
Of course, since we still would like to have a well defined geometric
structure, let us plug in the above equation back into (\ref{eq:Bianchi Identity}),
resulting in the following constraint: 
\begin{equation}
T^{\mu\nu}\partial_{\mu}\chi\left(x^{\rho}\right)+g^{\mu\nu}\partial_{\mu}\Lambda\left(x^{\rho}\right)=0,
\end{equation}
after using local conservation of the EMT, eq. (\ref{eq:LocalConservation}).
Note that a trivial solution would be to consider both $\alpha$ and
$\Lambda$ being constants, as usual.

In order to investigate the Newtonian limit, we remind ourselves that
locally $\Lambda$-effects are negligible (this is an empirical statement)
and one should recover Newtonian gravity. Therefore, after foliating
our spacetime with constant time slices labeled by $t$, one should
recover: 
\begin{equation}
\chi\left(x_{0}^{\rho}\right)=\frac{8\pi G_{0}}{c_{0}^{4}},
\end{equation}
where $x_{0}^{\alpha}$ represent our local configuration on the spacetime.
Now, being conservative, we can use this as an \emph{ansatz} for $\chi\left(x^{\alpha}\right)$
as a whole, so that we postulate the following functional dependence:
\begin{equation}
\chi\left(x^{\alpha}\right)=\frac{8\pi G\left(x^{\alpha}\right)}{c^{4}\left(x^{\alpha}\right)},
\end{equation}
in which $G$ and $c$ are variables over time and space. Therefore,
the constraint equation above reduces to\footnote{One could also have considered a more general constraint coming
from Bianchi identity: $(8\pi G/c^{4})\nabla_{\mu}T^{\mu\nu}=T^{\mu\nu}\partial_{\mu}G\left(x,t\right)-(4G/c)T^{\mu\nu}\partial_{\mu}c\left(x,t\right)+(c^{4}/8\pi)g^{\mu\nu}\partial_{\mu}\Lambda\left(x,t\right)$,
without assuming right away the typical minimally coupled conservation
law to be held. This would imply a violation of the local conservation
laws that could be further explored in future works. I thank Jerome
Quintin for this remark.}:
\begin{align}
    \left[\frac{1}{G}\partial_{\mu}G\left(x^{\alpha}\right)-\frac{4}{c}\partial_{\mu}c\left(x^{\alpha}\right)\right]\frac{8\pi G\left(x^{\alpha}\right)}{c^{4}\left(x^{\alpha}\right)}T^{\mu\nu}\left(x^{\alpha}\right)+ \nonumber \\
+\left[\partial_{\mu}\Lambda\left(x^{\alpha}\right)\right]g^{\mu\nu}\left(x^{\alpha}\right)=0,\label{eq:General Constraint}
\end{align}
which tells us how $c,$ $G$ and $\Lambda$ can vary altogether without
violating Bianchi identity, therefore preserving the underlying
geometrical framework intact, while at the same time preserving the
same local conservation laws we are familiar with. The above equation
is referred to as \emph{the general constraint }(GC)\emph{.} The cosmological
version of it has already appeared in the literature \cite{avelino,belinchon}.
One could also derive the second order constraint after operating
with $\nabla^{\nu}$ on the equation above.

Some important comments about this constraint follows:
\begin{enumerate}
\item Note that the variations of $G$ and $c$ are directly correlated
with the local matter/energy distribution while the variation of $\Lambda$
is correlated to the local geometry. This highlights an intrinsic
difference between the variation of $c$ and $G$ in comparison with
the variation of $\Lambda$, which will result in interesting implications
for their dynamics. 
\item Once that we do not expect that $\Lambda$ should have much of an
influence for local physics given current observations, one could
locally expect to have the following constraint being satisfied:
\begin{equation}
T^{\mu\nu}\left(\partial_{\nu}G-\frac{4G}{c}\partial_{\nu}c\right)\simeq0,
\end{equation}
which implies:
\begin{equation}
G\left(x^{\alpha}\right)=\frac{G_{0}}{c_{0}^{4}}c^{4}\left(x^{\alpha}\right),\label{eq:G as function of c}
\end{equation}
assuming $G\left(x_{0}^{\alpha}\right)=G_{0}$ for $c\left(x_{0}^{\alpha}\right)=c_{0}$.
This ties together the variation of Newton's gravitational coupling
to the variation of the speed of light. Now, since we have $G/c^{4}$
appearing in Einstein's equations, one can see that it could be changed
to $G_{0}/c_{0}^{4}$ locally, recovering the physics we are familiar
with, as long as variations of $c$ are small. In other words, for
negligible variations of $\Lambda$ and $c$ locally, Lorentz symmetry
is effectively restored\footnote{By Lorentz symmetry we are referring to the local group of transformations
that keep the infinitesimal (when considering curved spacetimes) line
element, given by $ds^{2}=-c^{2}\left(t,\vec{x}\right)dt^{2}+d\vec{x}^{2}$,
invariant. One could find odd the fact $g_{00}$ is spacetime dependent,
but it is important to remember that Lorentz transformations are infinitesimal
ones: $dt=\gamma\left(dt'+v/c^{2}dx'\right),\,d\vec{x}=\gamma\left(d\vec{x}'+\vec{v}dt'\right),$
where $\gamma=\left(1-v^{2}/c^{2}\right)^{-1/2},$ so that the spacetime
dependence does not play a role directly in the transformations. On
the other hand, it means that for different spacetime points, the
speed associated to the Lorentz symmetry will be different. This is,
of course, a naive way to think about a generalization of the Lorentz
group, and hopefully we will be able to address this issue in future
work. }. 
\item Note that vacuum solutions, given by $T_{\mu\nu}=0,$ imply $\Lambda$
to be a spacetime constant, while, at the same time, dropping any
constraint concerning the variation of $G$ and $c$. Of course, after
imposing $G$ and $c$ to be constants as well, one can recover the
usual Minkowski, de Sitter and anti-de-Sitter cosmological spacetimes.
This is a relevant observation also for non-cosmological vacuum solutions,
\emph{e.g.}, Schwarzschild's metric, since it means that the speed
of light and the Newton's coupling could be varying in principle,
albeit this would imply a phenomenological approach once the dynamics
of $c$ and $G$ would have to be imposed by hand. 
\item In standard GR, Einstein's equations for a traceless EMT theory implies:
\begin{equation}
R+4\Lambda_{0}=0.
\end{equation}
Since $\Lambda_{0}$ is constant, this also implies that the curvature
would have to be constant, as it is the case for electromagnetism,
for instance. However, in the current formalism, we see that one could
actually have a conformal field theory (traceless EMT) without necessarily
having constant curvature, given the fact that the above equations
are generalized to:
\begin{equation}
R\left(x^{\rho}\right)+4\Lambda\left(x^{\rho}\right)=0.
\end{equation}

\end{enumerate}
In the next sections, we start to explore some immediate consequences
of the variation of $c,$ $G$ and $\Lambda$ having the general constraint
imposed at all times.

\section{Cosmological background solution\label{sec:discussions}}

After we have discussed about the most immediate dynamical implications
that one could expect to have from the general constraint, we can
now analyze the implications of having $G,$ $c$ and $\Lambda$ varying
in Einstein's equations. In this section, we will restrict ourselves
to the background cosmology. 

We start off with a FRW-like \emph{ansatz}\footnote{Spatial curvature can be thought of as embedded in the $\vec{x}$
coordinates. }: 
\begin{equation}
g_{\mu\nu}=\text{diag }\left(-c^{2}\left(t\right),a^{2}\left(t\right),a^{2}\left(t\right),a^{2}\left(t\right)\right).\label{eq:FRW like ansatz}
\end{equation}
The reader should now be concerned with the fact that we could reparametrize
the time coordinate so that the time dependence of the speed of light
would disappear from the above \emph{ansatz}. However, this would
be misleading. Let us understand why that is the case. 

When we observe the CMB to be isotropic (and homogeneous after assuming
the Copernican Principle), we can use the CMB to define a whole class
of coordinate systems, which differ from one another by the function
$c(t)$ in the \emph{ansatz} above. In standard cosmology, this function
is meaningless since the speed of light is constant and we are allowed
to consider time-reparametrizations, so we may as well fix it to be
equal to a constant (in particular to be equal to $1$ in natural
units). However, for our given framework in which the speed of light
is varying, even though we are still allowed to consider time-reparametrizations
(given the covariance of the model), it is not possible to find a
time coordinate in which the speed of light would not be varying (since
its variation is not a coordinate artifact). Thus, time-reparametrizations
can only hide the effect of a varying-$c$ in the metric, for example
by finding a new time coordinate, call it $t'$, such that $c(t)dt=c_{0}dt'$.
If one does that, we interpret it as finding a new frame, to be called
CMB \emph{coflowing} frame, which is the frame that makes homogeneity
and isotropy explicit, but hides the effects of a varying-$c$ in
the metric\footnote{The nomenclature comes in analogy to the terminology \emph{comoving},
which refers to a spatial reparametrization of the physical coordinates,
$x_{p},$ to comoving coordinates, $x_{c}$, by $dx_{p}=a\left(t\right)dx_{c}.$
The comoving spatial coordinates move along with the Hubble flow,
remaining fixed. }. We do not want to use this preferential frame in the same way that
we do not want to use an inhomogeneous metric to describe background
standard cosmology, since it would be very hard to make homogeneity
explicit again. Hence, for us, in a general sense, the speed of light
has changed with cosmological time, where the cosmological time is
set up by the CMB evolution (which correlates this time coordinate
directly to the CMB temperature) and we keep this time dependence in our metric \emph{ansatz} so that
the effects related to a varying-$c$ remain explicit everywhere else.

That being said, we consider a perfect fluid EMT defined by:
\begin{equation}
T^{\mu\nu}=\frac{1}{c^{2}}\left(\varepsilon+p\right)U^{\mu}U^{\nu}+pg^{\mu\nu},\label{eq:Perfect fluid EM tensor-1}
\end{equation}
where $\varepsilon$ is the energy density, $p$ is the pressure,
and $U^{\mu}$ is the $4$-velocity, satisfying $g_{\mu\nu}U^{\mu}U^{\nu}=-c^{2},$
and, of course, $c=c\left(t\right)$ in our work\footnote{This sort of generalization has been known as minimal coupling \cite{Magueijo1}.
Following the arguments above about local variations of $c$ being
negligible, this seems to be a good \emph{ansatz} for the moment. }. 

Then, considering a perfect fluid, Einstein's equations can be reduced
to generalized Friedmann equations given by: 
\begin{gather}
H^{2}\left(t\right)=\frac{8\pi G\left(t\right)}{3c^{2}\left(t\right)}\varepsilon\left(t\right)-\frac{\Lambda\left(t\right)c^{2}\left(t\right)}{3}-\frac{kc^{2}\left(t\right)}{a^{2}\left(t\right)}\label{eq:1st Friedmann equation}\\
\frac{\ddot{a}\left(t\right)}{a\left(t\right)}=-\frac{4\pi G\left(t\right)}{3c^{2}\left(t\right)}\left[3p\left(t\right)+\varepsilon\left(t\right)\right]+H\left(t\right)\frac{\dot{c}\left(t\right)}{c\left(t\right)}-\frac{\Lambda\left(t\right)c^{2}\left(t\right)}{3},\label{eq:2nd Friedmann Equation}
\end{gather}
where $H\left(t\right)=\dot{a}\left(t\right)/a\left(t\right)$ and
`` $\dot{}$ '' stands for time derivatives. These same equations
have already been considered in \cite{avelino,belinchon}. Again,
a few comments are worth mentioning here: 
\begin{enumerate}
\item The first Friedmann equation has the same form as the usual one, but now
with $G,$ $c$ and $\Lambda$ being time variables.
\item The second Friedmann equation has an extra term, proportional to $\dot{c}\left(t\right),$
that disappears for $c=$ constant. However, now it becomes clear
that the background can expand in an accelerated fashion by demanding
that:
\begin{align}
-\frac{4\pi G\left(t\right)}{3c^{2}\left(t\right)}\left[3p\left(t\right)+\varepsilon\left(t\right)\right]+H\left(t\right)\frac{\dot{c}\left(t\right)}{c\left(t\right)}-\nonumber\\
-\frac{\Lambda\left(t\right)c^{2}\left(t\right)}{3}>0,\label{eq:acceleration condition}
\end{align}
which implies that even if $\Lambda=0$ we still can recover acceleration
with non-exotic matter ($3p+\varepsilon>0)$ given that $\dot{c}\left(t\right)\neq0$. 
\item Usually, in standard cosmology, we have $a$, $p$ and $\varepsilon$
as variables together with three equations: an equation of state and
two Friedmann ones. Here, the situation is trickier: we have $c,$
$G$, $\Lambda,$ $\varepsilon,$ $p$ and $a$ as variables. However,
we have two Friedmann equations, another one from the general constraint,
and an equation of state, giving four equations total\footnote{I thank Renato Costa for pointing this out to me.}.
Therefore, it is expected to have some hypothesis imposed by hand
concerning the time dependence of a pair of the set $\left\{ G,c,\Lambda\right\} .$
Besides, in the same sense that the equation of state represents the
underlying thermodynamic fluid being considered, one should now expected
that some underlying theory could provide the hypothesis for the variation
of this set of variables. In particular, if $\Lambda=0$, one needs
only one more equation \cite{cflation}.
\item The second Friedmann equation has a curvature correction due to the
variation of the speed of light. If we had used the CMB coflowing
frame, that term would be gone and we would have the following equations
after considering $c(t)dt=c_{0}dt'$:
\begin{align}
\left(\frac{a'}{a}\right)^{2}=\frac{8\pi c_{0}^{2}G\left(t'\right)}{3c^{4}\left(t'\right)}\varepsilon\left(t'\right)-\frac{\Lambda\left(t'\right)}{3}-\frac{k}{a^{2}\left(t'\right)}\label{eq:1st Friedmann equation-1}\\
\hspace*{0.6cm}\frac{a''\left(t'\right)}{a\left(t'\right)}=-\frac{4\pi c_{0}^{2}G\left(t'\right)}{3c^{2}\left(t'\right)}\left[3p\left(t'\right)+\varepsilon\left(t'\right)\right]-\frac{\Lambda\left(t'\right)}{3},\label{eq:2nd Friedmann Equation-1}
\end{align}
where `` $'$ '' denotes derivatives w.r.t. $t'$. These equations
look the same as what we have in standard cosmlogy while allowing
$c,$ $G$ and $\Lambda$ to vary. 
\item Another implication is the following: if we are not
considering cosmic scales, we can, in fact, choose a local coflowing
coordinate system (as one would be doing when considering black hole
solutions or physics on Earth, for instance) that is nor isotropic
or homogeneous, that is, not connected to the CMB. However, these
(local) frames would consist in a subclass of all the possible frames
that could be considered in principle (since for a varying-$c$ theory,
coflowing frames define a preferred class of frames). Since the CMB
defines another class of frames, the ones which are homogeneous and
isotropic, having among them a single \textit{one} which is also coflowing,
that implies this local subclass of coflowing frames is not always
compatible with the class defined by the CMB in the sense presented
by Padmanabhan \cite{Padmanabhan}. He argues that it is odd the fact
that operationally the CMB defines a preferred rest frame that does
not seem to select any preferred class of frames in sub-cosmic level
physics, which would be key if someone was to solve Einstein's equations
exactly and then to consider the average of the metric on cosmological
scales. In our framework, since now we do have a distinction between
the CMB comoving frame class and the coflowing one, we see that the
CMB would actually select a particular subclass of local frames: the
ones that are not coflowing, but rather comoving, therefore addressing
the problem raised by Padmanabhan. Given this clarification, one should
be concerned to what sort of effect we might have been observing that
could be due to the incompatibility between all these frames, since
we would be locally solving Einstein's equations with coflowing frames\footnote{Since there is no reason to make explicit any cosmological variations
of the speed of light when solving Einstein's equations locally.} and later expecting that these solutions would recover a comoving
non-coflowing frame when averaged over large scales. Although cosmology
has been quite successful as a whole, we still have two big elephants
in the room: dark matter and dark energy. In the current framework,
if these effects are actually due to the spacetime variation of $G$
and $c$ at the perturbation level, it is clear that a frame redefinition could in principle
hide these effects. \\

\end{enumerate}
Finally, we can also derive the continuity equation, namely the $0-$th
component of $\nabla_{\mu}T^{\mu\nu}=0,$ which gives:
\begin{equation}
\dot{\varepsilon}+3H\left(p+\varepsilon\right)=0.
\end{equation}
This can also be derived after taking the time derivative of (\ref{eq:1st Friedmann equation})
and substituting it into (\ref{eq:2nd Friedmann Equation}), and then
imposing the general constraint.

\subsection{$\Lambda-$less acceleration}

From a UV perspective, it is very reasonable to consider that $G$
and $c$ would be varying in the very early universe. Therefore, we
consider the case in which $\Lambda=0$, and investigate some of consequences
of the variation of the other \emph{constants.} 

We have seen already that the general constraint implies for this
case:

\begin{equation}
G\left(t\right)=\frac{G_{0}}{c_{0}^{4}}c^{4}\left(t\right),\label{eq:G as function of c-1}
\end{equation}
after considering $G$ and $c$ to be homogeneous. Then, assuming
a constant equation of state, $\omega=p/\varepsilon,$ and considering
(\ref{eq:1st Friedmann equation}) into (\ref{eq:2nd Friedmann Equation})
for $k=0,$ we have:
\begin{equation}
\frac{\dot{c}}{c}>\frac{1+3\omega}{2}\frac{\dot{a}}{a}.
\end{equation}
If this condition is satisfied, we are guaranteed to have acceleration.
If, for instance, we seek for a polynomial solution for the scale
factor, $a\left(t\right)\sim t^{n},$ $n>1,$ this demands the following
dynamics for the speed of light $c\left(t\right)\sim t^{m},$ $m>\frac{1+3\omega}{2}n.$ 

Needless to say that it would be much better if we could provide a
dynamical theory in which $c$ and $G$ would have their variations
given by an equation of motion, instead of considering any sort of
hypothesis by hand. This will be presented in \cite{cflation}, where
an action prescription will be introduced and early universe puzzles
will be dealt with, providing an alternative paradigm to inflation.

\section{A natural screening mechanism, or why we observe a cosmological constant
\label{sec:tres}}

A screening mechanism basically works in a way that some degrees of
freedom of a model are not accessible for a particular physical scale,
they are ``hidden''. For instance, in the context of cosmology, two
mechanisms are well known, they are called \emph{chameleons }\cite{Chameleon}
and \emph{symmetrons} \cite{Symmetron}. In short, these mechanisms
rely on a field, typically scalar, that has its mass (chameleon) or
its effective potential symmetry (symmetron) being dependent on the
local matter density. The importance of such mechanisms comes from
the fact that most of scalar fields coming from fundamental theories,
\emph{e.g.} String Theory, would imply strong violations of the equivalence
principle. However, such violations have not been observed in the
Solar System \cite{etvos}. Therefore, if someone hopes that such
fundamental fields could be responsible for any cosmological observations,
as dark energy for instance, such mechanisms would play an important
role to prevent these violations. 

Within the proposal of this paper, it is also possible to find something
that resembles a screening mechanism using a \emph{bootstrap }approach
after considering the GC. From eq. (\ref{eq:General Constraint}),
we have at the background level: 
\begin{equation}
\left(\partial_{t}G-\frac{4G}{c}\partial_{t}c\right)\varepsilon-\frac{c^{4}}{8\pi}\partial_{t}\Lambda=0.
\end{equation}
Now, we can think about this equation in two different scales: when/where
$\varepsilon$ is high, and when/where $\varepsilon$ is low. Therefore,
we could split this equation relative to these two scales,

\begin{equation}
\begin{cases}
\dot{G}-\frac{4G}{c}\dot{c}=\varepsilon_{\text{high}}^{-1}\left(\frac{c^{4}}{8\pi}\dot{\Lambda}\right)\simeq0 & ,\quad\varepsilon\,\text{high}\\
\frac{c^{4}}{8\pi}\dot{\Lambda}=\varepsilon_{\text{low}}\left(\dot{G}-\frac{4G}{c}\dot{c}\right)\simeq0 & ,\quad\varepsilon\,\text{low}.
\end{cases}
\end{equation}
 Therefore, we observe that when considering scales in which the energy
density is high, two things happen: i) the variation of
$G$ is fixed to the variation of $c$, which reproduces what we have discussed
above for our local physics; ii) it relaxes the temporal variation
of $\Lambda,$ since now its time-dependence can be ``more'' arbitrary,
once it is suppressed by $\varepsilon_{high}^{-1}$. On the other
hand, when $\varepsilon$ is low, we also observe two things: i) the
variation of $c$ and $G$ are not so tied together; ii) more importantly,
because the variations of $G$ and $c$ for those scales are suppressed,
the constraint tells us that effectively $\dot{\Lambda}\sim0$, once
its variation is suppressed by $\varepsilon_{low}/c^{4}.$ 

Note that we have related to these scales a sense of time (\emph{when})
and space (\emph{where}). For the former, we can imagine comparisons
between the background energy density in the early univese and nowadays,
being the first much higher than the second, what could help explaining
why the variation of $\Lambda$ seems to be non-existent today, even
though could have been much more prominent in the early universe. 

Regarding the sense of space, although we are assuming a homogeneous
description, this can also be considered a $0$th order description
of inhomogeneous regions. In doing so, we take the average energy
density of these regions. If we keep ourselves to galactic scales
the energy density will be much higher than if we average over cosmological
scales\footnote{This is the case since our homogeneous description of these scales
takes a small volume (galactic) compared to a large one (cosmological)
in order to consider the average energy density.}. Thus, even though this is a first approximation, it already tells
us that local energy density might also suppress how $\Lambda$ might
be varying. Of course, a full perturbation theory is needed to be
conclusive. 

Hence, this embedded natural screening mechanism brings new perspective
to the fact that we observe a cosmological constant in cosmological
observations, and still leaves room for its local variations. This
can be of great interest concerning the recent observations of \cite{BOSS}
that have created tension for the $\Lambda_{0}$CDM model concerning
the constancy of $\Lambda_{0}$. 

A small comment is necessary at this point. The general constraint
is an equation as valid as the Einstein's equations, since in this
framework both equations follow from the requirement to have the Bianchi
identity being satisfied and preserving the standard local conservation
laws. Therefore, our dynamical analyses that result from the general
constraint is automatically incorporated in the dynamics that one
finds coming from Einstein's equations. In other words, it is legitimate
to derive preliminary conclusions after looking only to the constraint
given all the equations are consistent among themselves.

\section{Prospects\label{sec:Prospects}}

In some sense, the framework presented here had already been briefly
considered in the literature, but many points were not particularly
addressed. Now that we have revived the model and shown some different
consequences of it, we have laid the ground for further work regarding
its own limitations and applications.

At the formalism level, we have emphasized the necessity of having
the dynamics of the fundamental constants being given by an action
principle/equation of motion. This is being developed \cite{cflation}
specifically in the framework discussed here, in which we consider
the background cosmology for $\Lambda=0.$ Having an action in someone's
hands, this opens room for considering different dynamics for the
fields associated to $c$ and $G$ by choosing different potentials,
in a similar fashion to what is done for inflation. This also allows
considering early universe puzzles and their resolutions without making
use of an \emph{ad hoc} field, such as the inflaton, since there is
a field associated to $c$ and $G$ (they are ``the same'' given the
general constraint for $\Lambda=0)$ comes for free. Evidently, this
can also be applied for late time cosmology, in particular investigating
dark energy. 

Another clear direction of work is to understand the cosmological
perturbations in this approach. We know that given that the equations
are self-consistent, we can consider a back of the envelope calculation
just looking at the general constraint at linear order in perturbations:
\begin{align}
\delta T_{\mu\nu}\left(\partial^{\mu}G-\frac{4G}{c}\partial^{\mu}c\right)+ \nonumber\\
+T_{\mu\nu}\left(\partial^{\mu}\delta G-\frac{4G}{c}\partial^{\mu}\delta c+\frac{4G}{c^{2}}\partial^{\mu}c\delta c-\frac{4}{c}\partial^{\mu}c\delta G\right)+ \nonumber\\
+\frac{c^{4}}{8\pi}g_{\mu\nu}\left(\partial^{\mu}\delta\Lambda+\frac{4}{c}\partial^{\mu}\Lambda\delta c\right)+\frac{c^{4}}{8\pi}\delta^{\mu}\Lambda\delta g_{\mu\nu}=0.
\end{align}
Here it is clear that variations of $c$ and $G$ are connected to
perturbations in the local matter density. Although we need to have
a proper treatment after defining gauge invariant variables, naively
it is expected that effects similar to cosmic accelerated expansion
could be a local effect due to such variations.

\section{Conclusions\label{sec:Conclusions}}

We have started the construction of a framework in which the fundamental
constants, $G,$ $c$ and $\Lambda,$ can be spacetime variables,
as long as their variations are constrained in relation to the local
geometry and local stress tensor. This framework does not provide
a dynamics for the variation of these constants, which leaves room
for an upgrade in which an action and equation of motion for these
variables could also be obtained, instead of having their dynamics
imposed by hand. When we have the full framework built, this will
provide a self-consistent and covariant approach for treating the
early universe puzzles after considering fundamental constants as
variables. 

Regardless, we can already appreciate within this formalism an explanation
of why we observe a cosmological constant today, realizing cosmic
accelerations by considering variations of the speed of light and
a better understanding of the interpretation of having a varying-$c$
theory which is also covariant. We also have presented extensive discussions
of the implications of choosing the CMB class of frames without assuming
the constancy of the fundamental constants in it. We hope that future
works will be able to bring other applications of the full framework,
not only at the background level, but for the perturbations as well.

\section{Acknowledgements }

I thank Renato Costa, Robert Brandenberger, Elisa Ferreira, Rodrigo
Cuzinatto, Jerome Quintin, Ryo Namba, Hossein Bazrafshan and Jonas
Pedro Pereira for useful discussions and for reading parts of the
manuscript. My research is supported by CNPq through the Science without
Borders (SwB).

\end{document}